\documentclass[pra,twocolumn,aps,longbibliography]{revtex4-2}
 \usepackage[english]{babel}
\usepackage{amsmath}
\usepackage{amssymb}
\usepackage{todonotes}
\usepackage[unicode=true,colorlinks=true,citecolor=blue,urlcolor=blue]{hyperref}
\usepackage{float}
\usepackage{bm}
\usepackage{setspace}
\usepackage{epsfig}
\usepackage{lipsum}
\usepackage{physics}
\usepackage{comment}
\renewcommand {\phi}{{\varphi}}
\newcommand {\rmi}{{\rm i}}
\newcommand {\rmd}{{\rm d}}

\newcommand {\e}{{\rm e}}

\newcommand{\gf}{\gamma_{\to}}
\newcommand{\gb}{\gamma_{\leftarrow}}

\renewcommand{\addcontentsline}[3]{}%

\makeatother

\begin{document}

\title{Bound, antibound and resonance  two-photon states in chiral waveguide QED}
\author{Jiaming Shi and Alexander Poddubny}

\email{poddubny@weizmann.ac.il}
\affiliation{Department of Physics of Complex Systems, Weizmann Institute of Science, Rehovot 7610001, Israel}

\begin{abstract}
We present a theoretical study of the two-particle spectrum $\omega(K)$ for the chiral waveguide QED setup of an array of two-level atoms directionally interacting with photons propagating along the waveguide. We demonstrate that for each pair center-of-mass momentum $K$  there exist distinct solutions with $\Im\omega\le 0$ in the two-particle spectrum, corresponding to bound, antibound and resonance states, in addition to the continuum of scattering states. Contrary to previous studies, which showed the bound and resonance-state spectra only over a limited range of $K$, the calculated spectrum is consistent across all $K$ values. An interesting finding is that the real part of the spectrum $\Re \omega(K)$ in the chiral model is gapless.  The calculated  dispersion law $\omega(K)$ provides  an effective model for the bound photon pairs also in a finite-size array, manifesting the topological non-Hermitian skin effect.
\end{abstract}

\date{\today}
\maketitle

\section{Introduction}\label{sec:intro}
Formation of bound, antibound and resonance states is an essential feature of the two interacting particles~\cite{moiseyev2011non}. Depending on the setup, the particle motion with respect to each other can be either fully localized (bound state) or can have a non-vanishing tail, describing a finite probability of dissociation (resonance state) or diverge (antibound state). In this work, we study  formation of such states in the chiral waveguide quantum electrodynamics (QED) setup~\cite{Sheremet2023}, for a periodic array of atoms chirally interacting with photons in the waveguide. 

Two essential features of the chiral waveguide QED model make it fundamentally interesting and different from other systems with interacting particles. The first feature is the nontrivial dispersion law of the model's elementary excitations, the polaritons. Polaritonic modes are formed by a hybridization of the two-level atom resonance at the frequency  $\omega_0$ and the propagating light with the $\omega=ck$ dispersion. The avoided crossings at $ck=\omega_{0}$, resulting from such a hybridization, render the polariton dispersion law strongly nonparabolic: the dispersion curves exhibit regions with slow and fast group velocities and also feature polariton band gaps. 
The second important feature of the chiral waveguide QED model is, by definition, the directionality of the photon radiation. The rates of the spontaneous photon emission by the atom to the left and to the right are different, $\gamma_\to\ne \gamma_\leftarrow$. The chiral waveguide QED setup can be realized when a transverse magnetic field is applied to the system~\cite{Lodahl2017,Lodahl2018,Prasad2020}, or when the couplings between the atom and the waveguide are nonlinear~\cite{Joshi_2023},  breaking the optical reciprocity.
\begin{figure}[b]
\centering
\includegraphics[width=0.48\textwidth]{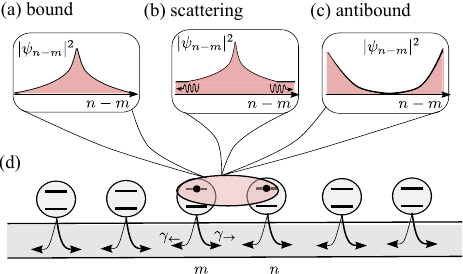}
\caption{Schematic illustration of the array of atoms chirally coupled to the waveguide (c) and the possible bound and  resonance states of two excitations (a,b). }\label{fig:schematic}
\end{figure}

The polariton dispersion law strongly depends on the dimensionless phase  $\varphi=\omega_0d/c$  that controls the ratio between the light wavelength at the atomic resonance $2\pi c/\omega_0$ and the array period $d$. In the limit, when the spacing between the atoms is small with respect to the light wavelength, $\varphi\ll 1$, and the emission is only in forward direction, $\xi=\gamma_\to/\gamma_\leftarrow\ll 1$, the system behaves as an effective homogeneous quantum nonlinear medium. The waveguide setup can then be directly mapped to another well-studied model of photons propagating in a cloud of  Rydberg atoms~\cite{Bienias2014,Calajo2022,Drori_2023,Das_2025}.  
Both the waveguide QED model with $\varphi\ll 1$ and the Rydberg polariton model feature a single two-photon bound state with unidirectional propagation. For finite values of $\varphi$ and $\xi$, the waveguide QED model becomes more complicated. Generally, there are at least two branches of bound states as well as resonance states~\cite{Zhang2019arXiv}.

 Despite a significant number of recent theoretical studies in the literature, there's apparently no complete understanding of the bound and resonance state spectra in the chiral waveguide QED setup ~\cite{Zhang2019arXiv,Poddubny2019quasiflat,bakkensen2021,Calajo2022,Schrinski_2022,
jen2021bound,Levy_Yeyati_2025,
Shi_2025}. Even the basic question of the number of bound and resonance states for a general value of the polariton center-of-mass wave vector $K$ is not fully clear. Many studies, including some of our works, restricted the consideration to a certain range of $K$ and/or ignored resonance states~\cite{Poddubny2019quasiflat,Shi_2025}. In the very comprehensive recent works \cite{bakkensen2021,Schrinski_2022}, the resonance states were calculated only for certain values of $K$ and disappeared for other values of $K$. This raises questions about the spectrum's continuity and general analytical properties.  Antibound solutions, where the frequency is real and the amplitude increases with the distance between the particles, have not been reported in the WQED setup to the best of our knowledge.   The goal of the current work is to provide a more complete picture, accounting for all possible complex-$\omega$ solutions for arbitrary values of $K$, $\varphi$, and $\xi=\gamma_\leftarrow/\gamma_\to$.

We demonstrate that for general values of $0<\varphi<\pi $ and $0<\xi<1$, in addition to the usual continuum of scattering states, there exist distinct non-increasing-in-time nontrivial solutions for the relative motion of the photon pair for each center-of-mass wavevector $K$ with complex eigenfrequencies $\omega(K)$, $\Im\omega(K)\le 0$. These solutions encompass bound, antibound, and resonance states. The real part of the spectrum turns out to be gapless as long as the structure is chiral: for any $\widetilde\omega$ there exists a real value $ K$  so that $\Re \omega(K)=\widetilde\omega$.  In addition to the numerical calculations, we present analytical answers, describing all four solutions for the limiting values of $K\to 0,\pm \omega_0/c,2\pi\pm \omega_0/c,2\pi,\ldots$

 Our  results provide rigorous foundation behind the formation of edge-localized bound states in the finite chiral structure.  In our previous work \cite{Shi_2025}, it has been found that the bound photon states concentrate at the edge of the finite chiral structure, which is the essence of the non-Hermitian skin effect (NHSE)~\cite{Yao2018, Bergholtz2021}.  This has been interpreted as a result of directional dissociation of bound photon pairs inside the continuum --- the pairs can propagate to the left, while those propagating to the right dissociate.  Instead of directly solving for the resonance states, we proposed in Ref.~\cite{Shi_2025} a phenomenological dispersion model inside the continuum to explain this effect. Now we can use rigorous resonant-state dispersion, rather than a phenomenological model. This allows us, in particular, to  understand the puzzling finding of Ref.~\cite{Shi_2025}, that for atoms preferentially emitting to the right, $\gamma_\to>\gamma_\leftarrow$, the bound state is concentrated at the left edge, and vice versa.

 The rest of the paper is organized as follows. In Sec.~\ref{sec:model} we outline our calculation approach for the relative motion of the two photons. We generalize the method from Ref.~\cite{Poddubny2019quasiflat} and show that the infinite-range coupling chiral waveguide QED model is in fact a tight-binding model with second-nearest-neighbor couplings in disguise. The formation of bound and scattering states is then akin to edge-state formation in the usual solid-state setup. Next,  in Sec.~\ref{sec:results} we discuss the results of numerical and analytical calculations. We start with recalling known results for a fully chiral case\cite{Calajo2022} in Sec.~\ref{sec:chiral} and then consider  an arbitrary value of the chirality parameter in Sec.~\ref{sec:gen}.  Section~\ref{sec:NHSE} discusses finite-size array and provides the connection of our results to the non-Hermitian skin effect. The main results are summarized in Sec.~\ref{sec:summary}.
 Auxiliary derivations are given in the Appendices.

\section{From long-range coupling to tight binding model}\label{sec:model}
We start by writing the effective non-Hermitian Hamiltonian of the structure~\cite{Sheremet2023,Poddubny2024skin}:
\begin{equation}\label{eq:H0}
H_{\rm eff}=\omega_0\sum_{m=1}^N\sigma_m^\dag\sigma_m^{\vphantom{\dag}}+ \sum_{m,n=1}^NH_{m,n}\sigma_m^\dag\sigma_n^{\vphantom{\dag}}\:,
\end{equation} 
where 
\begin{equation}
\label{eq:H}
H_{m,n}=-\rmi\begin{cases}
 \gf \e^{\rmi\varphi |m-n|}, &m>n\:,\\
 \frac{\gf+\gb}{2}, &m=n\:,\\
 \gb \e^{\rmi\varphi |m-n|}, &m<n\:.
\end{cases}
\end{equation}
 Here, $\omega_0$ is the atom resonant frequency, and  $ \gf=2\gamma_{\rm 1D}/(1+\xi)$  and  $ \gb=2\gamma_{\rm 1D}\xi/(1+\xi)$, 
are the rates of spontaneous emission of right- and left-propagating waveguide photons. The directionality of the emission is quantified by the parameter $\xi$; the non-chiral case corresponds to $\xi=1$. 
The Hamiltonian
Eq.~\eqref{eq:H} is written in the Markovian approximation with traced-out photon modes, which is valid for $\gamma_{\rm 1D}\ll \omega_0$. 

In this work, we restrict ourselves to the propagation of two excitations in the periodic structure that has translational symmetry. The center-of-mass wave vector $K$ of the two excitations is then a good quantum number. The two-photon wave function $|\psi\rangle = \sum_{m,n} \psi_{mn}\sigma_m^\dag \sigma_n^\dag|0\rangle $ can be sought in the form
\begin{equation}\label{eq:K}
    \psi_{mn} = \e^{\rmi K (m+n)/2-2\rmi\omega_0t} \chi_{m-n}\:,
\end{equation}
with $\chi_{m-n}=\chi_{n-m}$. The boundary condition is $\chi_{0}=0$: a single atom can not host two excitations simultaneously. This can also be seen as a short-range repulsion.
Substituting the ansatz Eq.~\eqref{eq:K} into Eq.~\eqref{eq:H0}, we obtain the following Hamiltonian~\cite{Calajo2022}:
\begin{align}\label{eq:4}
        &\hat{H}_K =\sum_{r,r'>0} 
        [{H}_K]_{rr'} \sigma_r^\dag\sigma_{r'}^{\vphantom{\dagger}},
        \:{H}_K=\gamma_\to F_\to+\gamma_\leftarrow F_\leftarrow\:,
        \\
        &
        [F_{\to/\leftarrow}]_{rr'}=-\rmi (\e^{\rmi \varphi_{\to/\leftarrow}|r-r'|}+\e^{\rmi \varphi_{\to/\leftarrow}(r+r')})\:,\nonumber
\end{align}
for the interacting photon pairs, where $\varphi_{\to/\leftarrow}=\varphi\mp K/2$.
Crucially, the Hamiltonian Eq.~\eqref{eq:4}  features a long-ranged waveguide-mediated coupling, inherited from the original Hamiltonian Eq.~\eqref{eq:H}. While the Hamiltonian \eqref{eq:4} still remains tractable due to the presence of translational symmetry~\cite{bakkensen2021,Schrinski_2022}, the dense structure of the matrix $[\hat{H}_K]_{rr'}$ hinders the analysis of the complex spectrum. 

It has been noted in Ref.~\cite{Poddubny2019quasiflat}, that the inverse of the dense matrix $F_0(\varphi)=-\rmi\e^{\rmi\varphi |r-r'|}$, that has the same structure as the first term in $F_{\to/\leftarrow}$, is a tri-diagonal matrix:
\begin{equation}
[F_{0}^{-1}]_{r,r\pm 1} = \frac{1}{2\sin \varphi}\,,\quad 
[F_{0}^{-1}]_{r,r} = -\cot \varphi\:,\,\qquad
\end{equation}
with the only complex elements at the edges,
\begin{equation}
[F_{0}^{-1}]_{1,1} =
[F_{0}^{-1}]_{N,N} =
\frac{\rmi-\cot\varphi}{2}\:.
\end{equation}
The $F_{\to/\leftarrow}$ matrices differ from $F_0(\varphi\mp K/2)$ only by a rank-1 term $\e^{\rmi(\varphi\mp K/2)(r+r')}$. Hence, their inverse can be readily calculated by the Sherman-Morrison formula \cite{Bartlett1951}. The result is still a tridiagonal matrix
\begin{equation}\label{eq:iF}
[F_{\to/\leftarrow}^{-1}]_{r,r'} =
\begin{cases}[F_{0}^{-1}(\varphi_{\to/\leftarrow})]_{rr'}, & r{~\text{or} }~r'\ne 1,N\:,\\
-\cot 2\varphi_{\to/\leftarrow}, & r=r'=1\:,\\
\frac{\rmi}{2}-\frac1{2}\cot \varphi_{\to/\leftarrow}, & r=r'=N\:,\\
\end{cases}
\end{equation}
that differs from $F_0$ only at the edges. 

Multiplying the  equation
$H_K\chi=2\omega\chi$ with $H_K$ from Eq.~\eqref{eq:4} by 
$F^{-1}_\to$ and $F^{-1}_\leftarrow$ we find 
\begin{equation}\label{eq:final}
\widetilde H\chi\equiv 
     (2\omega F^{-1}_\to F^{-1}_\leftarrow-\gamma^{\vphantom{-1}}_\to F^{-1}_\leftarrow-\gamma^{\vphantom{-1}}_\leftarrow F^{-1}_\to) \chi=0\:.
\end{equation}
Equation~\eqref{eq:final} is a generalized sparse eigenvalue problem for $\omega$. The matrix in the left-hand-side is a tridiagonal one, the matrix $F^{-1}_\to F^{-1}_\leftarrow$ is a pentadiagonal one. Thus, the problem has, at most, next-nearest-neighbor couplings. 
The bulk solution can be sought in the form satisfying Bloch's theorem 
\begin{equation}\label{eq:Bloch}
\chi_r\propto z^r\:,
\end{equation}
where $z\equiv \exp(\rmi \kappa)$ is a propagation constant and $\kappa$ is the (complex) wave vector describing the relative motion of the two particles. For $|z|<1$,  $\kappa$ has a nonzero imaginary part, which corresponds to the inverse localization length for the relative motion, that is, the inverse size of the bound state. 

The dispersion equation then assumes the usual form 
\begin{equation}\label{eq:dispersion}
D(z,\omega)=\sum\limits_{r=-2}^2z^r t_r=0\:,
\end{equation}
for the tight-binding problem, with 
\begin{align}
t_{\pm 2}&=\frac{\omega}{2\sin\varphi_\to \sin\varphi_\leftarrow}\:,\\
t_{\pm 1}&=-\frac{\gamma_\to}{2\sin\varphi_\leftarrow}-\frac{\gamma_\leftarrow}{2\sin\varphi_\to}-\omega\left(\frac{\cot\phi_\leftarrow}{\sin\phi_\rightarrow}+\frac{\cot\phi_\rightarrow}{\sin\phi_\leftarrow}\right)\:,\nonumber\\
t_0&=\gamma_\to \cot\varphi_\leftarrow+
\gamma_\leftarrow \cot\varphi_\to
\nonumber\\\nonumber&+2\omega  \cot \varphi_\to
\cot \varphi_\leftarrow+\frac{\omega}{\sin\varphi_\to\sin\varphi_\leftarrow}\nonumber\:.
\end{align}
This equation is equivalent to those in \cite{bakkensen2021,Schrinski_2022}, but the direct correspondence to the second-nearest-neighbor problem makes it a bit more transparent. For example, it is now trivial to observe that there exist two pairs of solutions with $z=z_a,1/z_a,z_b,1/z_b$ for each $\omega$.

Such radical reduction from the full problem Eq.~\eqref{eq:4} makes the analysis conceptually simpler and very straightforward.
The model has energy shifts at the sites $1$, $N$, as well as the modified matrix elements between the sites $1,2$ and $N,N-1$. Thus, the bound and resonance state formation can be seen as a result of these boundary defects. 
In particular, the diagonal matrix element $t_0$ acquires the shift
\begin{multline}
    \delta t_0\equiv \widetilde H_{1,1}-\widetilde H_{2,2} = -\frac{\gamma_\rightarrow}{\sin 2\phi_\leftarrow}
               - \frac{\gamma_\leftarrow}{\sin 2\phi_\to}
               \\+ 2\omega\Bigl(\cot 2\phi_\to\cot 2\phi_\leftarrow
               - \cot\phi_\to\cot\phi_\leftarrow
               \\- \frac{1}{4\sin\phi_\to\sin\phi_\leftarrow}\Bigr)\:.
\end{multline}
The tunneling matrix elements get edge corrections
\begin{align}
    \delta t_1\equiv \widetilde H_{1,2}-\widetilde H_{2,3}=\frac{\omega }{\sin 2\varphi_\to\sin\varphi_\leftarrow},\\
    \delta t_{-1}=\widetilde H_{2,1}-\widetilde H_{3,2}=\frac{\omega }{\sin \varphi_\to\sin2\varphi_\leftarrow}\:,
\end{align}
and the boundary equations for Eq.~\eqref{eq:final} become
\begin{align}
r=1:\quad & (t_0+\delta t_0)\,\chi_1 + (t_1+\delta t_1)\,\chi_2 + t_2\,\chi_3 = 0\,,\label{eq:bc1}\\
r=2:\quad & (t_1+\delta t_{-1})\,\chi_1 + t_0\chi_2 + t_1\,\chi_3 + t_2\,\chi_4 = 0\,.\nonumber
\end{align}
For $|z_a|, |z_b|\le 1$ we can look for the bound state in the form 
\begin{equation}\label{eq:psi}
\chi_n = A\,z_a^n + B\,z_b^n\,,\quad n\geq 1\,,
\end{equation}
and the boundary conditions read 
\begin{equation}\label{eq:bound_2x2}
\begin{pmatrix}
z_a\,f_1(z_a) & z_b\,f_1(z_b) \\
z_a^2\,f_2(z_a) & z_b^2\,f_2(z_b)
\end{pmatrix}
\begin{pmatrix} A \\ B \end{pmatrix} = 0\,,
\end{equation}
where 
\begin{align}
f_1(z) &= t_0+\delta t_0 + (t_1+\delta t_1) z + t_2 z^2\,,\label{eq:f1}\\
f_2(z) &= (t_1+\delta t_{-1})/z + t_0 + t_1z + t_2 z^2\,.\nonumber
\end{align}
A nontrivial solution exists when the determinant of this 2x2 system vanishes:
\begin{equation}\label{eq:bound_det}
{z_b\,f_1(z_a)\,f_2(z_b) = z_a\,f_1(z_b)\,f_2(z_a)\,.}
\end{equation}
Equation \eqref{eq:bound_det}, together with the dispersion Eq.~\eqref{eq:dispersion}, presents a closed system of equations to find the bound states that should be solved numerically. Resonance states are found from the same system but in the complex $\omega$ plane.

The main complication for numerical solution is that there exist various choices for $z_a$, $z_b$ for every $\omega$, nontrivially governed by Eq.~\eqref{eq:dispersion}.
However, the frequency $\omega$ can be eliminated from these equations to yield a single closed-form 8-degree polynomial for $z$. The roots of this polynomial can be readily found and  filtered  to keep only the physical solutions. Our code for numerical calculations is publicly available \cite{poddubny2026topobound_figures}.

The advantage of our tight-binding formulation over the one in Refs.~\cite{bakkensen2021,Schrinski_2022} is that the number of roots is fixed, and there is no need to solve complex-valued transcendental equations --- all the equations are polynomial ones. We have also verified the bound-state and some resonance- solutions against the direct numerical diagonalization of both Eqs.~\eqref{eq:4} and Eqs.~\eqref{eq:final} on a truncated basis.

\begin{figure*}[t]
\centering
\includegraphics[width=\textwidth]{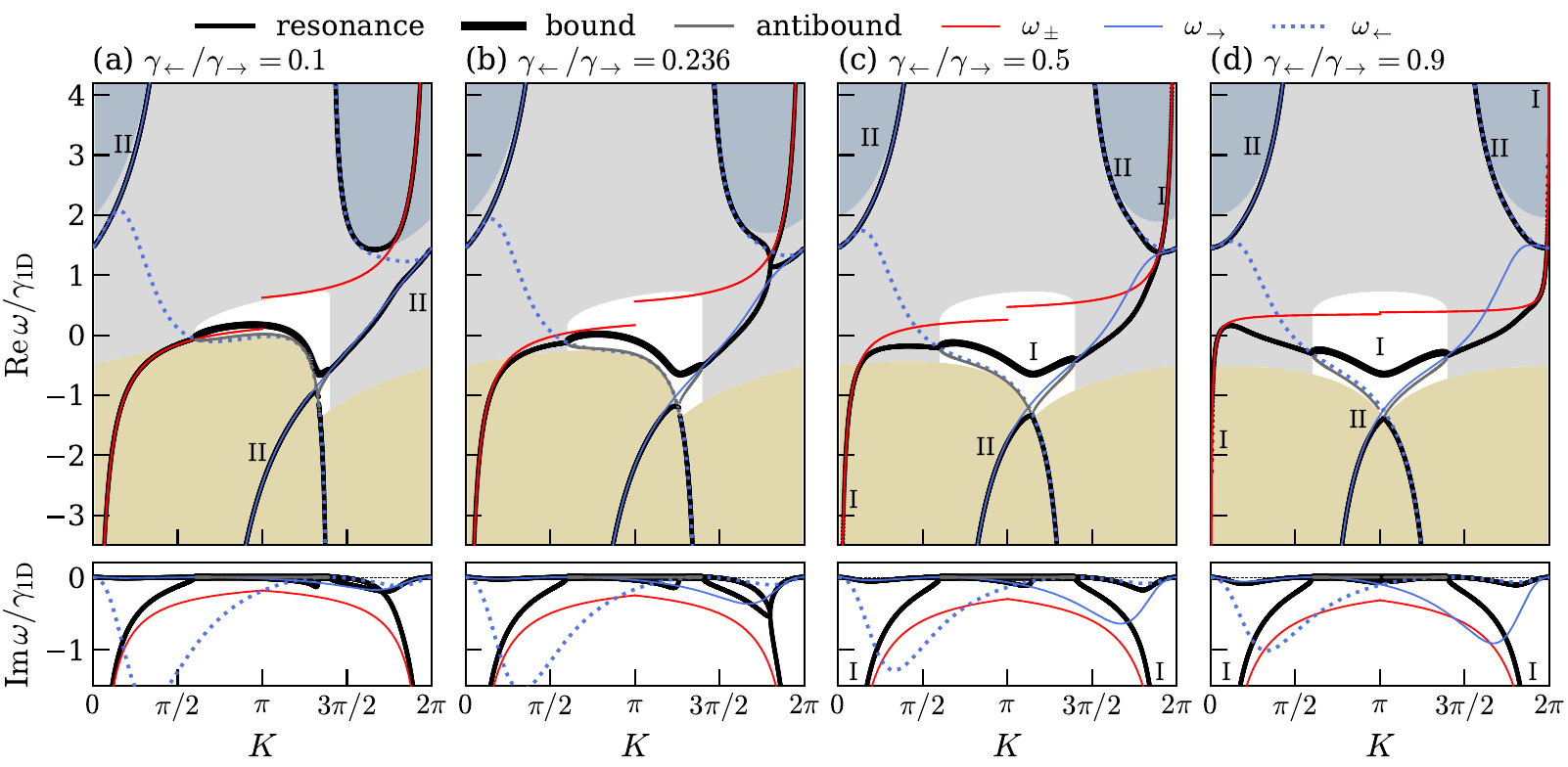}
\caption{\label{fig:spectrum_re_im}
Complex two-photon bound-, antibound- and resonance state spectrum versus center-of-mass momentum $K$
for four values of the chirality parameter $\gb/\gf$ at fixed array period $\varphi\equiv \omega_0d/c=0.3\pi$. Thin and thick parts of the black curves correspond to resonance and bound states, respectively.  Shaded blue, gray and yellow regions mark the two-photon continuum associated with two upper-band, upper and lower-band and two lower-band polaritons; the white
area is the gap. Analytic asymptotes are shown for $\omega_\pm$ (thin
red solid, Eq.~\eqref{eq:asympt1}), $\omega_\to$ (thin blue solid, Eq.~\eqref{eq:asympt2}), and $\omega_\leftarrow$
(thin blue dotted). Only solutions with $\Im \omega\le 0$ are shown, the spectrum is mirror-symmetric for $\Im \omega>0$ .}
\end{figure*}

\section{Bound, antibound and resonance states dispersion}\label{sec:results}
\subsection{Fully chiral case}\label{sec:chiral}
In this section, we will show how the general formalism can be used to reproduce the known answer~\cite{Calajo2022}
\begin{equation}
   \omega(K)=\gamma_\to \cot \phi_\to\equiv -\gamma_\to\cot \left(\frac{K}{2}-\varphi\right)\label{eq:bound-chiral}
\end{equation}
for the bound state frequency in the fully chiral system, when $\gamma_\to=2\gamma_{\rm 1D}$, $\gamma_\leftarrow=0$.
In this case, only the $F_\to$ term survives in Eq.~\eqref{eq:4},
\begin{equation}
{H}^K = \gamma_\to F_\to\,,
\end{equation}
and the generalized eigenvalue problem Eq.~\eqref{eq:4} simplifies to the standard eigenvalue problem
\begin{equation}\label{eq:chiral}
F_\to^{-1}\chi = \frac{\gamma_\to}{2\omega}\chi\:,
\end{equation}
This is equivalent to a simple nearest-neighbor tight-binding chain with defects at the edges.
The bulk dispersion equation for the wave Eq.~\eqref{eq:Bloch} becomes just
\begin{equation}\label{eq:disp-chiral}
    \frac{1}{2\sin \varphi_\to}(z+1/z)=\frac{\gamma_\to}{2\omega}+\cot\varphi_\to\:.
\end{equation}
At $n=1$ instead of Eq.~\eqref{eq:bc1} we find
\begin{equation}\label{eq:bc3}
    \frac{1}{2\sin \varphi_\to}z=\frac{\gamma_\to}{2\omega}+\cot2\varphi_\to\:.
\end{equation}
Subtracting Eq.~\eqref{eq:bc3} from Eq.~\eqref{eq:chiral} we find $z=\cos\varphi_\to$.
Substituting this back to Eq.~\eqref{eq:bc3} we recover  Eq.~\eqref{eq:bound-chiral} for the chiral bound state dispersion.

\begin{figure*}[t]
\centering
\includegraphics[width=\textwidth]{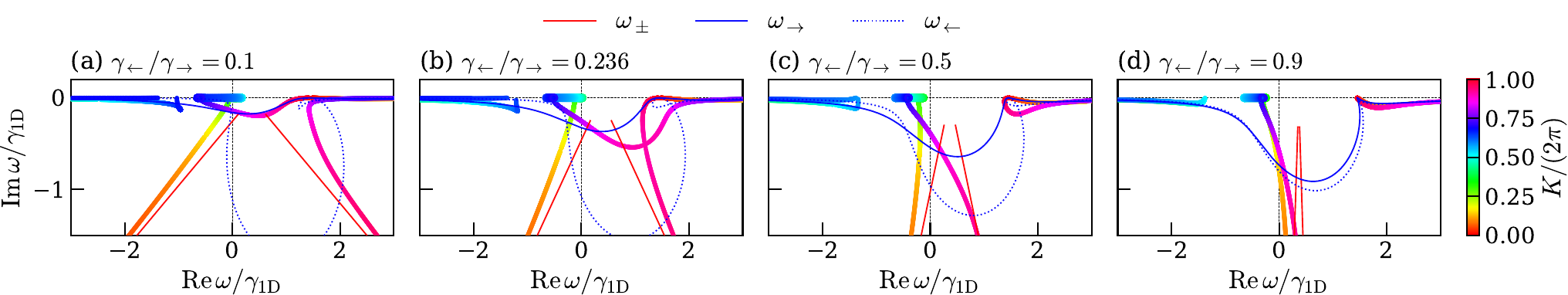}
\caption{\label{fig:spectrum_complex}
Spectrum $\omega(K)$ in the plane of complex $\omega$ 
parametrized by $K=0\ldots 2\pi$. Calculation parameters are the  same parameters as in  Fig.~\ref{fig:spectrum_re_im}. Thin red, blue solid, and blue dotted curves are the
analytic asymptotes $\omega_\pm$, Eq.~\eqref{eq:asympt1}, and $\omega_{\to/\leftarrow}$, Eq.~\eqref{eq:asympt2}. Antibound solutions are not plotted. Only solutions with $\Im \omega\le 0$ are shown, the spectrum is mirror-symmetric for $\Im \omega>0$.
}
\end{figure*}

\subsection{General case}\label{sec:gen}
\begin{figure*}[t]
\centering
\includegraphics[width=\textwidth]{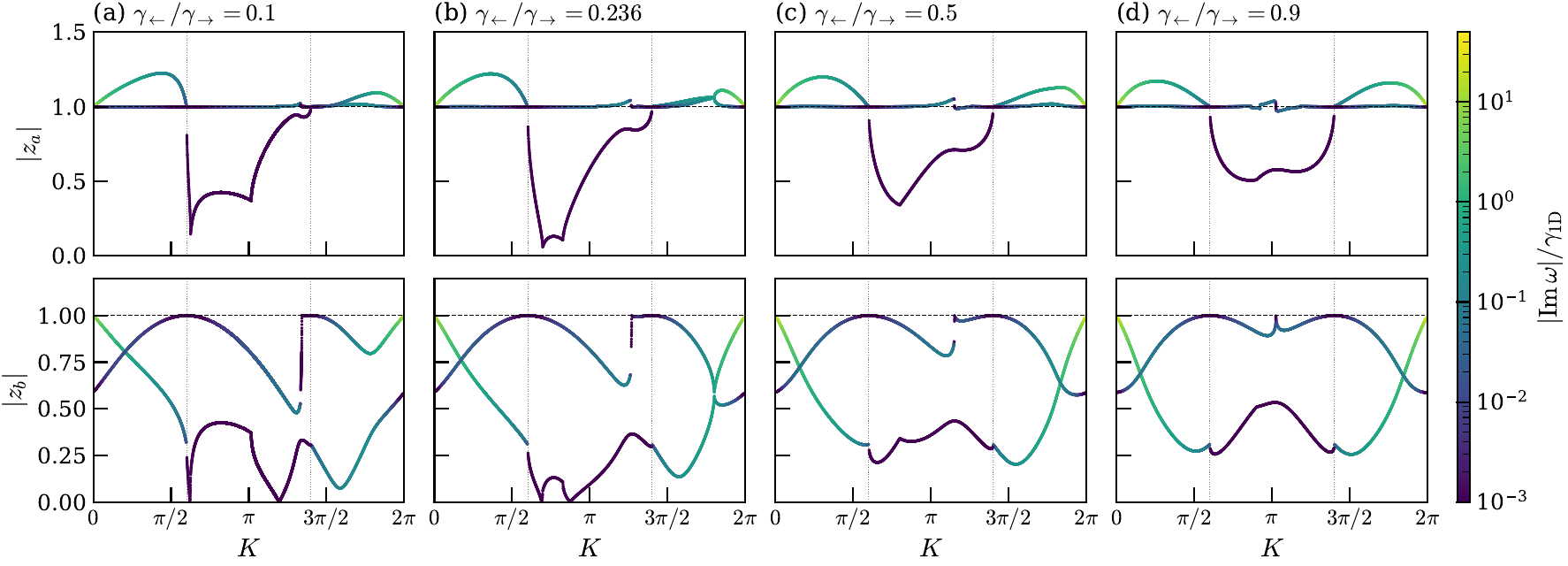}
\caption{\label{fig:z_moduli}
Absolute values of the $z_{a}$ (top) and $z_{b}$ (bottom)  coefficients, describing the two-photon state spatial profile Eq.~\eqref{eq:psi} depending on the 
center-of-mass momentum $K$. The parameters are the same as in Fig.~\ref{fig:spectrum_re_im} and
Fig.~\ref{fig:spectrum_complex}.
Only the roots with $\Im\omega\le 0$ are shown, antibound solutions are not plotted. Marker
color encodes the value of $|\Im\omega|$ on a logarithmic scale.}
\end{figure*}

We now proceed to the results of general numerical calculation. Figure~\ref{fig:spectrum_re_im} presents the real (top row) and imaginary (bottom row) parts of the dispersion curves $\omega(K)$ for four values of the chirality parameter $\gamma_\leftarrow/\gamma_\to$. The columns are ordered so that the degree of chirality decreases from left to right.
Figure~\ref{fig:spectrum_complex} presents the same dispersion curves but in the plane of complex $\omega$, parametrized by the wavevector $K=0\ldots 2\pi$. The absolute values of the solutions $z_a$ and $z_b$ from Eq.~\eqref{eq:psi} for every state are shown in Fig.~\ref{fig:z_moduli}.

Interestingly, contrary to the $F_{0}^{-1}$ matrices, the edge elements
$[F_{\to/\leftarrow}^{-1}]_{r,r'}$ for $r=r'=1$ in Eq.~\eqref{eq:iF} are real. The only complex matrix elements are those at $r=r'=N$. Since we are considering the limit $N\to \infty$, the equation we solve  does not involve these matrix elements and deals only with real-valued $F_{\to/\leftarrow}^{-1}$. Hence, it has an $\omega\leftrightarrow \omega^{*}$ symmetry and produces both decaying in time solutions with $\Im \omega<0$ and increasing in time with $\Im \omega>0$, as well as the bound states with $\Im \omega=0$. These solutions should be distinguished by their behavior at $r\to \infty$ or time $t\to \infty$, similarly as it is done for outgoing and incoming waves $G(z)\propto \exp(\pm \rmi \omega |z|/c)$ when solving for a Green function $G(z)$ of the Helmholtz equation with a point source $G''(z)+(\omega/c)^{2}G(z)=-\delta(z)$. For a finite value of $N$ the imaginary contribution $\rmi /2$
in $[F_{\to/\leftarrow}^{-1}]_{N,N}$ breaks the symmetry between $\omega$ and $\omega^{*}$ and enforces  the decaying-in-time solutions. In our approach with infinite $N$, we chose to show only the decaying-in-time branches with $\Im \omega\le 0$ out of each pair $\omega,\omega^{*}$.

Thin curves in Fig.~\ref{fig:spectrum_complex} show the analytical asymptotes discussed below.
The shaded areas correspond to the continua of scattering states. Since the single-polariton dispersion has two branches, an upper one and a lower one, for two-polariton states, one can distinguish between the scattering states where both polaritons are in the upper band (blue area), in the lower band (yellow area), or one in each band (gray area).

For low chirality, column (d), the dispersion features a true bound state in the gap, $2\varphi<K<2\pi-2\varphi$. This solution has both $|z_a|,|z_b|<1$, so the relative motion amplitude Eq.~\eqref{eq:psi} decays exponentially with the distance between the particles.
When the bound states branch enters the continuum, it becomes a resonance state, and $|\Im \omega|$ diverges when $K$ approaches $0$ or $2\pi$ (these bound and resonance states are marked by ``I'' in Fig.~\ref{fig:spectrum_re_im}c--d).   Such diverging solutions are well seen in the lower row of Fig.~\ref{fig:spectrum_re_im} and are well approximated by the analytical thin red curves
\begin{equation}
    \omega_{\pm}=\frac{\gamma_\leftarrow-\gamma_\to\pm 2\rmi\sqrt{\gamma_\to\gamma_\leftarrow}}{2K}+\frac{\gamma_{\rm 1D}}{2}\cot\varphi\:,\label{eq:asympt1}
\end{equation}
(and similar for $K\to 2\pi$).
This equation is derived in the Appendix~\ref{sec:app:A}. Interestingly, as long as the structure has some nonzero chirality degree, $\gamma_\to\ne \gamma_\leftarrow$, the real part of this branch is gapless. The value of $\Re \omega$ spans the whole range  $-\infty$ to $+\infty$ as $K$ varies. This can be also seen directly from Eq.~\eqref{eq:asympt1} where the leading term is $\Re\omega_{\pm}\propto (\gamma_\leftarrow-\gamma_\to)/K$.

 In addition to the bound state, there also exists a resonance state with $\Re \omega>0$ and $\Im \omega<0$ for $K\approx 0,2\pi$. The resonance state is seen just below the upper (dark blue shaded) continuum of scattering states~\cite{Zhang2019arXiv} and is marked by numbers ``II'' in Fig.~\ref{fig:spectrum_re_im}c--d. At exactly one  point $K=0,2\pi$ the decay rate of this resonance state tends to zero, and it becomes a true bound state~\cite{Calajo2022}.  We note, that the amplitude of the relative motion wavefunction $\chi_{n}$, corresponding to the resonance state, is actually increasing with the distance between the particles: $\Im \omega<0$ implies $|z|>1$,
 so $\chi_{n}\propto |z|^{n}$ diverges. Such non-normalizable behavior is typical for a complex-frequency resonance state~\cite{moiseyev2011non,Ching_1998}. 
 
 At $K=2\varphi, 2\pi-2\varphi$, the resonance states dispersion branches have diverging $\Re \omega$. They can be approximated by  the asymptotic expressions for $\omega_{\to}(K)$,
 $\omega_{\leftarrow}(K)$:
\begin{gather}
\omega_{\to} = -\gamma_{\to}\cot\varphi_{\to} + \Sigma_{\to}\:,\nonumber\\
\Sigma_{\to} = \frac{-\rmi\gamma_{\leftarrow}(1 - \e^{2\rmi\varphi_{\leftarrow}}\cos 2\varphi_{\to})}{2(1 - \cos\varphi_{\to}\e^{\rmi\varphi_{\leftarrow}})^2}\,,
    \label{eq:asympt2}
\end{gather}
and similarly for $\omega_\leftarrow(K)$, see Appendix~\ref{sec:app:B} for the derivation. The asymptotic expressions are shown by thin solid and dotted blue curves in Figs.~\ref{fig:spectrum_re_im},\ref{fig:spectrum_complex}. In particular, this means that  $\Re \omega\to +\infty$ for $K=2\varphi-0$, then emerges from the $-\infty$ for $K\to 2\varphi+0$. The upper- and lower- frequency resonance states thus form the same dispersion branch, connected at $\omega\to \infty$.

{In the range $2\varphi<K<2\pi-2\varphi$, there also exists one more  solution of the edge equations Eq.~\eqref{eq:bound_det} with real $\omega$ and $|z_a|>1$, shown by dark gray lines in Figure~\ref{fig:spectrum_re_im}. Due to  $|z_a|>1$ the amplitude of this  ``antibound'' solution increases with the distance between particles.
 At the band gap edges  $K=2\varphi,2\pi-2\varphi$, the eigenvalue corresponding to the antibound solution coalesces with that of the bound state branch. An exceptional point is formed as a result. This exceptional point leads to  small square-root-like features in $\Re \omega(k)$ that are best seen for the branch I of Fig.~\ref{fig:spectrum_re_im}(d) at  $K\to 2\varphi,2\pi-2\varphi$.} To avoid the clutter, we show the antibound state branch only in Fig.~\ref{fig:spectrum_re_im}.

\begin{figure}[b]
\centering
\includegraphics[width=\columnwidth]{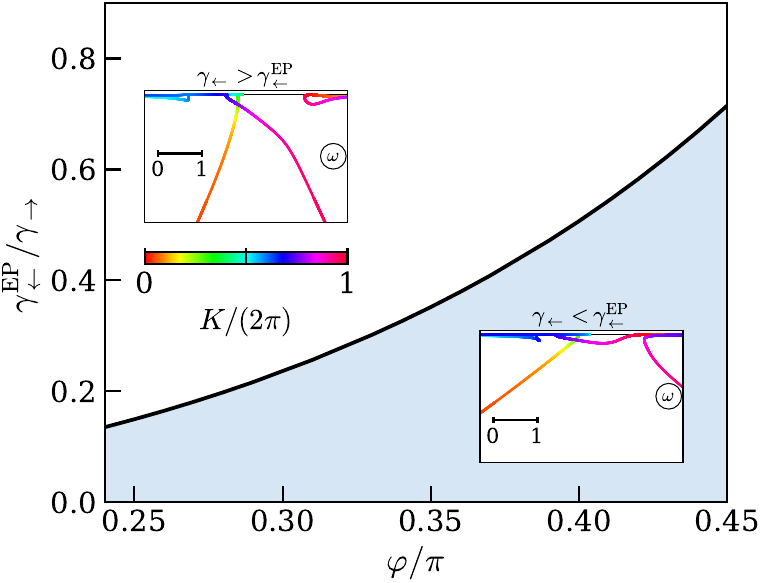}
\caption{\label{fig:EP_vs_phi}
Exceptional-point position $\gb^{\rm EP}/\gf$ depending on the phase 
$\varphi/\pi$. The shaded region below the curve corresponds to the regime of strong chirality. Insets show schematically show dispersion contours $\omega(K)$ for $\gamma_{\leftarrow}/\gamma_{\to}=\{0.4,0.1\}$ and $\varphi=0.3\pi$, similarly to Fig.~\ref{fig:spectrum_complex}.}
\end{figure}

As the system becomes more chiral, the dispersion curves remain qualitatively the same until $\gamma_\leftarrow/\gamma_\to\equiv \gamma_\leftarrow^{{\rm EP}}/\gamma_\to\approx 0.236$ (column b). The two branches then merge, forming an exceptional point (EP) at $K\approx 1.8\pi$ and then reconnect in a different way, compare Fig.~\ref{fig:spectrum_complex}(a--c). For strong chirality, there appears a gap for $\Re \omega$ at $2\pi-\varphi<K<2\pi$. Importantly, the real part of the dispersion for resonance and bound states still remains gapless. The value of $\gamma_\leftarrow/\gamma_\rightarrow$, when this exceptional point is formed, can be seen as a transition between the strong- and weak-chirality regimes.  The dependence of $\gamma_\leftarrow^{{\rm EP}}/\gamma_\to$  on $\varphi$ is shown in Fig.~\ref{fig:EP_vs_phi}. As $\varphi$ decreases, the 
ratio $\gamma_\leftarrow/\gamma_\rightarrow$ has to be smaller in order to get into the strong-chirality regime. The possible interpretation could be that a smaller $\varphi$ implies a larger band gap in the continuum spectrum $\propto 1/\sin\varphi $~\cite{Poddubny2019quasiflat}. For a larger gap, it becomes harder to achieve mixing between different polariton branches, which may be required for the formation of an exceptional point.

The binding energy for the bound state in the gap decreases for stronger chirality. The bound state branch approaches the lower continuum and ceases to exist at $\gamma_{\leftarrow}\to 0$. At the same time,  for $\gamma_\leftarrow\to 0$, the resonance state branches marked by numbers II in Fig.~\ref{fig:spectrum_re_im} merge and form a single chiral bound state  with the dispersion
\eqref{eq:bound-chiral}.

\section{NHSE for the resonant states}\label{sec:NHSE}
We now turn to the analysis of the non-Hermitian skin effect (NHSE) --- concentration of the bound pairs at the edge of an array with a finite number of atoms. 
Our numerical calculations, reported in Ref.~\cite{Shi_2025}, demonstrated that for $\gamma_{\to}>\gamma_{\leftarrow}$, that is, photons emitting mostly to the right, the bound pairs concentrate at the left edge of the  finite array, and vice versa.  This has been linked to the negative group velocity of the bound states dispersion branch, but still remains quite counterintuitive. Here we provide a more rigorous analysis based on the complex resonant-state dispersion, branches $\omega_{\rm I}(K)$, that labeled by ``I'' in the top row of Fig.~\ref{fig:spectrum_re_im}(c,d). 

We introduce the  effective Hamiltonian for the motion of the center of mass of the bound pair: 
\begin{align}\label{eq:Heff}
\mathcal H_{m,n}^{(\rm eff)}&=\int_{\mathcal C} \frac{\rmd K}{2\pi} \e^{\rmi K(m-n)}  \omega_{\rm I}(K),\\\nonumber
\mathcal C&=[\delta_0\ldots 2\varphi-\delta_1]\cup [2\varphi+\delta_1\ldots 2\pi-2\varphi-\delta_1]\\\nonumber&\cup [2\pi-2\varphi+\delta_1\ldots 2\pi-\delta_0]\:. \nonumber
\end{align}
{Such a Hamiltonian can be defined only when $\gamma_\leftarrow>\gamma_\leftarrow^{\rm EP}$, that is, when the chirality is relatively weak and the bound state of interest corresponds to a single dispersion branch $ \omega_{\rm I}(K)$.}
The integration interval $\mathcal C$ is just the Brillouin zone $[0\ldots 2\pi)$ excluding the intervals around the singularities. In our numerical calculations we used $\delta_0 = 0.0001, \delta_1 = 0.002$. {The results for the bound state branch in the spectral range of interest, $|\Re \omega|\sim\gamma_{\rm 1D}$, seem to be weakly depending on the regularization. The value of $\delta_0$ affects only the eigenstates of $\mathcal H^{(\rm eff)}$ with large energy that are beyond this effective description.} 

\begin{figure}[t!]
\centering
\includegraphics[width=\columnwidth]{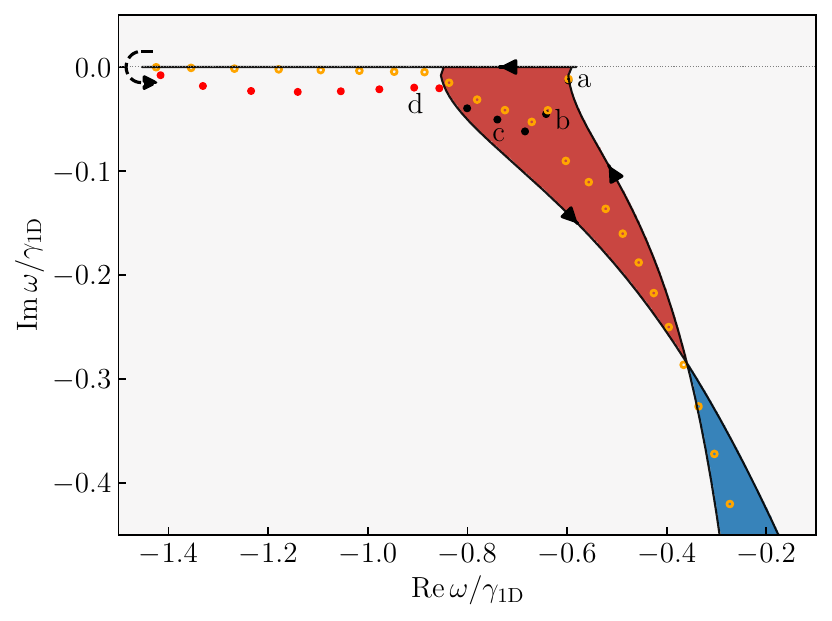}
\caption{\label{fig:OBC}
Complex energy spectrum of the {bound and resonant states}. The solid curve presents the dispersion contour $\omega_{\rm I}(K)$ {under periodic boundary conditions.} Regions characterized by distinct winding numbers relative to the dispersion contour are color-coded: red ($w=+1$), white ($w=0$), and blue ($w=-1$). Arrows show the direction of integration along the $K$-path for the winding number. The red and black solid symbols show the spectrum of the bulk and edge-localized {bound states under open boundary conditions.} 
Labels (a--d) indicate the states whose spatial profile is shown in Fig.~\ref{fig:modes}.
The open orange symbols show the eigenstates of the effective Hamiltonian Eq.~\eqref{eq:Heff}. The calculation has been performed for $\varphi= 0.35\pi$, $N=40$, and $\xi=0.7$. 
}
\end{figure}

\begin{figure}[t!]
\centering
\includegraphics[width=\columnwidth]{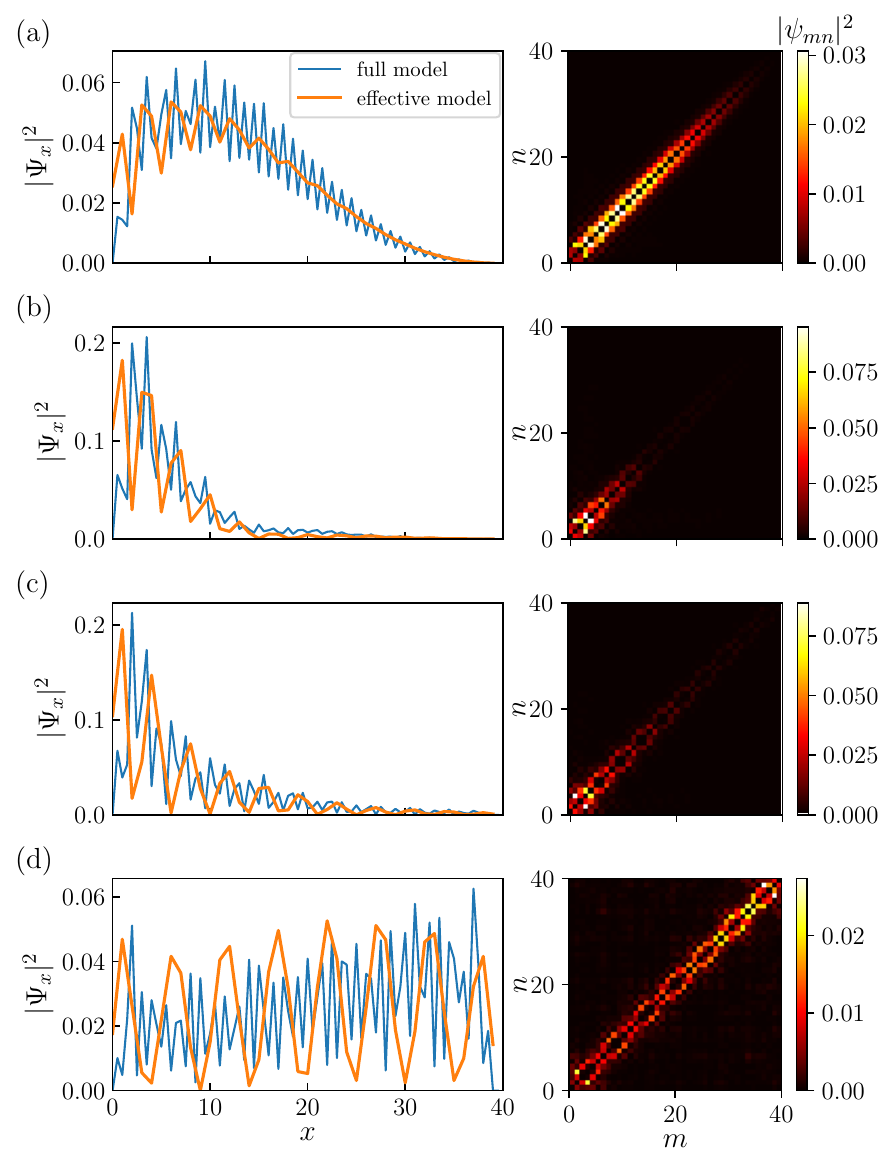}
\caption{\label{fig:modes}  Spatial profile of the bound pairs and the corresponding eigenstates of the effective Hamiltonian Eq.~\eqref{eq:Heff} in a finite-size array with $N=40$ sites. The complex energies of the states for each row (a-d) are indicated in Fig.~\ref{fig:OBC}. The left panels present the distribution of the center-of-mass of the bound pairs (thin blue curves) and the eigenstates of $\mathcal H^{(\rm eff)}$ (thick orange curves). The right panels present the  distributions of the  bound states depending on the coordinates of both excitations $m$ and $n$. 
}
\end{figure}

Figure~\ref{fig:OBC} shows the energy spectrum under periodic and open boundary conditions, and 
Fig.~\ref{fig:modes} shows the bound eigenstates in a finite-size array.
 The solid curves in Fig.~\ref{fig:OBC}  present the dispersion contour $\omega_{\rm I}(K)$. The solid and open symbols show the spectrum of the bound states in the finite-size array, found from the exact diagonalization of the  Hamiltonian \eqref{eq:H} and of the effective center-of-mass Hamiltonian Eq.~\eqref{eq:Heff}. The results demonstrate a good agreement for the complex energies {of the bound states and the nearby eigenvalues of $\mathcal H^{(\rm eff)}$. } {The winding number at the reference point $E_b$ in the complex energy plane is defined as $w =  \int_0^{2\pi} \arg\!\left[H(k) - E_b \right] \rmd k/(2\pi)$. }
 
Moreover, Fig.~\ref{fig:modes}   well captures the spatial distribution of center of mass of the bound states.  This can be seen from comparison of {orange and blue} curves in the {left column} of Fig.~\ref{fig:modes}, illustrating the spatial profile of several characteristic eigenstates indicated in Fig.~\ref{fig:OBC}. The right column of Fig.~\ref{fig:modes} presents the full two-photon joint probability
$|\psi_{mn}|^2$.  {The center-of-mass distribution is defined as $|\Psi_x|^2 = \sum_{m} |\psi_{m,2x-m}|^2$, representing the total probability density along the diagonal.}

Both effective model and full model succesfully reproduce edge localization of the bound states, whose energies lie inside the loop of the complex branch $\omega_{\rm I}(K)$ in Fig.~\ref{fig:OBC}.  The main difference is the presence of additional oscillations in the full model. The oscillations stem from the  fine structure of the bound state relative motion wavefunction $\chi_{n-m}$. They  do not affect the position of the center-of-mass of the bound pair on a larger spatial scale. {Another difference is that the wavefunction in the effective model is nonzero at the edges of the structure, while the actual bound state population decreases to zero. This discrepancy can be explained by the finite size of the bound state which prevents the center of mass to be exactly at the edge. Such effect is present  for nonchiral arrays as well~\cite{Poddubny2019quasiflat}. Similarly boundary suppression is also known in the physics of semiconductor nanostructures for excitons, that is, bound electron-hole pairs. For example, the resonant contribution to the polarization of the medium, associated with the exciton, vanishes at some finite distance from the edge of the interface (so-called ``dead layer'')~\cite{Excitons,IvchenkoPikus}.}

The overall satisfactory agreement of the effective  model with the full one is somewhat surprising, given that the effective model Eq.~\eqref{eq:Heff} ignores other dispersion branches and the continuum of the scattering states. In a finite-size array the center-of-mass momentum is no longer a good quantum number. Indeed, for weaker localized bound states the agreement becomes worse, see Fig.~\ref{fig:OBC}. { The finite-size effects including an admixture of the other modes could be the explanation.} 

Importantly, the part of the dispersion contour $\omega_{\rm I}(K)$, encircling the {edge-localized} bound-state spectrum in the finite array, has a winding number of {$+1$}, see the arrows in Fig.~\ref{fig:OBC}. According to \cite{Zhang_2020}, this means that the finite-size eigenmodes are concentrated at the left edge of the array. This agrees with the numerical calculations in Fig.~\ref{fig:modes}{(a-c)} and explains why the bound states are concentrated the left edge despite the atoms emitting preferentially to the right. 

The description based on the winding number does not contradict the ``negative group velocity'' description in our previous work Ref.~\cite{Shi_2025} but rather complements it. 
According to that previous interpretation, the bound state group velocity in the ``unidirectional region'', where $\omega_{\rm I}(K)$ is real and $\omega_{\rm I}(2\pi-K)$ is complex, is negative, $\rmd \omega_{\rm I}/\rmd K<0$, hence bound states concentrate at the left edge. At the same time, the negative group velocity means that the bound state dispersion branch is skewed so that 
$\omega_{\rm I}(K=-2\varphi)>\omega_{\rm I}(K=2\pi-2\varphi)$, see   Fig.~\ref{fig:spectrum_re_im}(a--d).
If the group velocity were positive, the skew sign would flip, $\omega_{\rm I}(K=-2\varphi)<\omega_{\rm I}(K=2\pi-2\varphi)$. The two complex-energy  parts of the branch $\omega_{\rm I}(K)$ would then not cross each other as they do now in Fig.~\ref{fig:OBC}. The inner loop in Fig.~\ref{fig:OBC} would disappear, the winding number sign would change, and the bound states would concentrate at the right edge, contrary to what we observe now.

\section{Summary and Outlook}\label{sec:summary}
To summarize, we present numerical and analytical results for the bound, antibound and resonance states dispersion in the chiral waveguide QED problem. The main difference from previous studies~\cite{bakkensen2021,Schrinski_2022} is that we were able to obtain the dispersion of resonance states $\omega(K)$ for the complex frequencies $\omega$ and real $K$. This allowed us to continuously track the complex  $\omega(K)$  dispersion branches in the whole range of real $0<K<2\pi$ with the only exception being the polariton dispersion singularities $K=0,\pm 2\omega_{0}/c\equiv \pm 2\varphi$, where $\omega(K)$ diverges. The equations for the relative motion have been rewritten as an effective tight-binding model with second-nearest-neighbor coupling and edge defects. Such a simplified formulation ensures that no solutions are lost. We have also obtained the analytical asymptotic approximations to the numerical results.
An interesting finding is that as long as the system has any chirality, the real part of the complex spectrum becomes gapless, that is, for every real $\widetilde\omega$ there exists a real value of $K$ so that $\Re \omega(K)=\widetilde\omega$. This spectral feature reminds of the edge states in topological insulators, which span the gap between the continua of propagating states~\cite{Hasan2010}, so are also inherently gapless. It is, however, unclear whether there is any deeper meaning behind this connection. The considered two-photon solutions arise solely from edge defects in the effective tight-binding model, rather than from the topologically nontrivial bulk dispersion of photon pairs.

We have connected the derived resonant-state dispersion to the topological non-Hermitian skin effect (NHSE), that is, edge-concentrated bound photon pairs, predicted in our previous work~\cite{Shi_2025}. While Ref.~\cite{Shi_2025} relied on a phenomenological resonant-state dispersion, here we were able to use rigorous numerical results. In particular, we now understand that the bound pairs are concentrated at the left edge when the photons are mostly emitted to the right (and vice versa) because of the sign of the winding number of the complex resonant-state dispersion. Moreover, we present an effective Hamiltonian for the bound state in a finite system, based on the Fourier transform of the corresponding dispersion branch $\omega(K)$. The one-dimensional model for the quantized motion of the center-of-mass of the bound pair in a finite structure quantitatively reproduces both the energies and the spatial distribution of the bound states in a finite array. 

It is interesting whether these results can be generalized to more complicated waveguide QED setups, for example, when atoms are coupled not only to waveguide photons but also to free-space photons, or to waveguide photons with a more complex dispersion law. There is hope for at least an approximate generalization in the vicinity of the edges of the Brillouin zone, which is inspired by the previous results for an atom chain in free space~\cite{Zhang2020d}.

\acknowledgements
This work has been stimulated by multiple discussions with Tom\'{a}s Levy-Yeyati, Mengjie Yang  and Alejandro Gonz\'{a}lez-Tudela.
It  has been supported by research grants from the  Minerva Foundation, from the Center for New Scientists, and the Center for Scientific Excellence at the Weizmann Institute of Science.

\appendix
\section{Derivation of Eq.~\eqref{eq:asympt1}}\label{sec:app:A}
In order to derive Eq.~\eqref{eq:asympt1}, we first note that the values of the propagation constants $z$ for the two solutions merge at $K\to 0,2\pi$. This can be seen directly from the top row of Fig.~\ref{fig:z_moduli}, or it can be proved from the equation Eq.~\eqref{eq:bound_det}.  Since both $z_{a}$ and $z_{b}$ satisfy the dispersion equation Eq.~\eqref{eq:dispersion}, $D(z_{a,b})=0$, we can rewrite Eq.~\eqref{eq:bound_det} as 
\begin{equation}\label{eq:bound_det2}
z_b\widetilde f_1(z_a) \widetilde  f_2(z_b) = z_a \widetilde  f_1(z_b) \widetilde  f_2(z_a)\:,
\end{equation}
with $\widetilde f_{1,2}(z)\equiv f_{1,2}(z)-D(z)$.  For $K=0$ Eq.~\eqref{eq:bound_det2} can be then simplified to
\begin{equation}\label{eq:bound_det2b}
\frac{\omega^{2}}{z_{a}^{2}z_{b}^{2}}(z_{a}-z_{b})(z_{a}z_{b}-1)(z_{a}-\cos\varphi)(z_{b}-\cos\varphi)=0\:.
\end{equation}
The roots we are interested in at $K\to 0$, $\omega\to \infty$,  correspond to $z_{a}=z_{b}=\cos\varphi$.

Next, we look for the solutions  of the dispersion equation  in the asymptotic form 
\begin{equation}\label{eq:K0}
w\equiv z+\frac1{z}=2\cos\varphi+uK,\: \omega=\frac{\Omega}{K}\:.
\end{equation}
The dispersion equation  Eq.~\eqref{eq:dispersion} can be rewritten  as 
\begin{equation}
t_{2}(w^{2}-2)+t_{1}w+t_{0}=0\:,
\end{equation}
which is a quadratic equation for $u$. Its discriminant is 
$4\Omega^2 + 4\Omega (\gamma_{\leftarrow}-\gamma_{\rightarrow}) +4\gamma_{\rm 1D}^2$.
Since $z_{a}=z_{b}$, the two roots coalesce, and the discriminant has to be zero. This results in
\begin{equation}
\Omega=\Omega_{\pm}=\frac{\gamma_{\to}-\gamma_{\leftarrow}}{2}\pm\rmi \sqrt{\gamma_{\to}\gamma_{\leftarrow}}\:,
\end{equation}
which yields the leading in $1/K$ term in Eq.~\eqref{eq:asympt1}.

The subleading  correction term $\gamma_{\rm 1D}\cot \varphi/2$ in Eq.~\eqref{eq:asympt1} requires a bit more involved algebra. We verified numerically and analytically that it does not depend on the ratio $\gamma_{\leftarrow}/\gamma_{\to}$, but only on their sum. We present below the derivation only for $\gamma_{\leftarrow}=\gamma_{\to}=\gamma_{\rm 1D}$. The solution of the dispersion equation can be written in the general form
\begin{multline}
\omega(w,K)=\\\frac{w\gamma_{\to}\sin\phi_{\leftarrow}-2\gamma_{\to} \cos\phi_{\to}\sin\phi_{\leftarrow}+(``\to''\leftrightarrow ``\leftarrow'')}{(w-2\cos\phi_{\to})(w-2\cos\phi_{\leftarrow})}\:.
\end{multline}
Now, at $K\to 0$, $\gamma_{\to}=\gamma_{\leftarrow}$, $w=z+1/z=2\cos\varphi$  both numerator and denominator diverge as $K^{2}$. Their ratio converges to 
\begin{equation}
\omega(w=2\cos\varphi,K\to 0)=\frac{\gamma_{\rm 1D}}{2}\cot\varphi\:,
\end{equation}
which is exactly the subleading term in Eq.~\eqref{eq:asympt1}. It is this term that has to be added to the expansion Eq.~\eqref{eq:K0} to provide the next-order correction in $1/K$.

\section{Derivation of Eq.~\eqref{eq:asympt2}}\label{sec:app:B}

In order to derive Eq.~\eqref{eq:asympt2} we start from the general Eq.~\eqref{eq:4},
\begin{equation}
    \gamma_\to F_\to \chi+\gamma_\leftarrow F_\leftarrow \chi=2\omega\chi\:.
\end{equation} 
For $\gamma_{\leftarrow}=0$, the solution is given by Eq.~\eqref{eq:bound-chiral}. Next, we use the first-order perturbation theory in $\gamma_\leftarrow$. We substitute the solution  $\chi_n=(\cos\phi_\to)^n$, obtained in Sec.~\ref{sec:chiral},
and write the correction term
\begin{equation}
\Sigma_\to=\gamma_\leftarrow\frac{\langle \chi_n|F_\leftarrow|\chi_n\rangle}{\langle \chi_n|\chi_n\rangle}\:.
\end{equation}
The sum in the numerator is 
\begin{equation}
\langle \chi_n|F_\leftarrow|\chi_n\rangle=\sum_{n,n'=1}^\infty 
(\beta^{|n-n'|}+\beta^{n-n'})\mu^n\mu^{n'}
\end{equation}
with $\beta=\exp(\rmi\phi_\leftarrow)$, $\mu=\cos\phi_{\to}$. The first term sums to 
\begin{equation}
\sum_{n,n'=1}^\infty 
\beta^{|n-n'|}\mu^{n}\mu^{n'}=\frac{\mu^2}{1-\mu^2}\frac{1+\mu\beta}{1-\mu\beta}\:,
\end{equation}
and the second term sum factorizes to $\sum_{n,n'=1}^\infty\beta^{n-n'}\mu^n\mu^{n'}=(\beta\mu)^2/(1-\beta\mu)^2$. The normalization sum is just $\langle \chi_n|\chi_n\rangle=\mu^2/(1-\mu^2)$. Gathering all together, we obtain  Eq.~\eqref{eq:asympt2} from the main text.

%

\end{document}